\documentclass[eqsecnum,aps]{revtex4}
\usepackage{amsmath,amssymb,amsthm,bm}
\usepackage{psfrag}
\usepackage[dvips]{graphicx}
\usepackage[utf8]{inputenc} 
\usepackage{hyperref}

\begin{document}
\title{Plane vortex pairs, colorful structures and low-lying Dirac modes}
\author{Seyed Mohsen Hosseini Nejad}
\email{smhosseininejad@semnan.ac.ir}
\affiliation{
Faculty of Physics, Semnan University, P.O. Box 35131-19111, Semnan, Iran}
   
\begin{abstract}

We investigate the influence of topological charges on non-stable zero and near-zero modes of the single uni-color plane vortex pairs. We combine the uni-color plane vortices with the spherical vortex and also construct the plane vortex pairs with two colorful vortices. The stability of zero and near-zero modes is analyzed where various boundary conditions for the fermions are checked. In addition, plane vortex pairs may show the role of the topological charges for changing the chirality of fermions. The results clearly indicate characteristic properties for spontaneous chiral symmetry breaking.  \\  \\ 
\end{abstract}


\maketitle

\section{INTRODUCTION}\label{Sect1}
Non-perturbative QCD is dominated by the
 phenomena of quark confinement and spontaneous chiral symmetry breaking ($\chi$SB). Center vortex ~\cite{tHooft:1977nqb,vinciarelli:1978kp,yoneya:1978dt,cornwall:1979hz,mack:1978rq,nielsen:1979xu} removal annihilates confinement and restores chiral symmetry~\cite{Faber:2017alm,deforcrand:1999ms,alexandrou:1999vx}. This result encourages that vortices are responsible for $\chi$SB. However, the physical mechanism for $\chi$SB through vortices is still unclear. Lattice simulations indicate that center vortices may be responsible for the topological charge ~\cite{Bertle:2001xd,Engelhardt:2000wc,Engelhardt:2010ft,Hollwieser:2010mj,
Hollwieser:2011uj,Schweigler:2012ae,Hollwieser:2012kb,Hollwieser:2014mxa,Hollwieser:2015koa,Altarawneh:2015bya,Hollwieser:2015qea,HosseiniNejad:2015oeu,HosseiniNejad:2018adw} and $\chi$SB~\cite{engelhardt:1999xw,reinhardt:2000ck,engelhardt:2002qs,leinweber:2006zq,bornyakov:2007fz,hollwieser:2008tq,bowman:2010zr,hollwieser:2013xja,brambilla:2014jmp,hoellwieser:2014isa,trewartha:2014ona,trewartha:2015nna,HosseiniNejad:2016hna}. According to the Banks-Casher relation \cite{banks:1979yr} a finite density of near-zero modes of the Dirac operator leads to the non-zero chiral condensate and therefore $\chi$SB. 

 The single pairs of the uni-color plane vortices have been introduced in Ref. \cite{Jordan:2007ff} where they do not have any topological charges. For the overlap Dirac operator in the background of these single uni-color vortex pairs, some zero and near-zero modes have been calculated which can be removed by appropriate boundary conditions. Therefore these modes are not stable. 

In this paper, we concentrate on the behavior of these non-stable zero and near-zero modes in the presence of the topological charges. In this way, we combine the uni-color plane vortices with the spherical
vortex and also construct the plane vortex pairs with two colorful vortices. We analyze the relation between center-vortex gauge field configurations with nontrivial topological charge and the presence of zero or near-zero modes of the Dirac operator in the background of the proposed gauge fields. The interplay between both properties is interesting as the topological charge is indeed related with the spectral properties of the Dirac operator. The construction of the spherical vortex has been
explained in~\cite{Jordan:2007ff,Hollwieser:2010mj,Schweigler:2012ae}. We present what occurs for the non-stable zero and near-zero modes of single pairs of the uni-color plane vortices after adding the spherical vortex with topological charge. The stability of these modes which are not due to the topological charges is analyzed in the background of the spherical vortex where various boundary conditions for the fermions are checked. Then, we investigate the zero and near-zero modes attracted by topological charge through constructing the plane vortex pairs with two colorful vortices. The construction of the plane vortices with one colorful plane has been introduced in \cite{HosseiniNejad:2015oeu} where one zero mode is calculated which agrees with the expectations from the Atiyah-Singer index theorem ~\cite{atiyah:1971rm,brown:1977bj,adams:2000rn}. In Ref.~\cite{hollwieser:2013xja}, the near-zero mode is calculated when there is a time distance between spherical vortex and anti-vortex. Now, we study the low-lying modes for plane vortex pair with two colorful vortices where there is a distance in spatial direction between two colorful plane vortices. We show how the topological charges of the two colorful vortices form near-zero modes from zero modes through interactions. The stability of zero and near-zero modes is checked through various boundary conditions for the fermions. In addition, comparing the Dirac modes of the uni-color and colorful plane vortex pairs, one may confirm the change in chirality of fermions by the topological charges. The results clearly present the role of the topological charges for creating stable zero and near-zero modes and show that center vortices precisely capture the nontrivial topological charge and the order parameter for chiral symmetry breaking. 
  
 In section~\ref{Sect2} we combine the plane vortices with the colorful regions. In the background of the proposed configurations we compute
the eigenmodes of the overlap Dirac operator. In subsection~\ref{Subsect1}, we combine the uni-color plane pairs with spherical vortex and the influence of topological charge is investigated on non-stable zero and near-zero modes of the uni-color plane pairs. The plane vortex pair with two colorful vortices is constructed in ~\ref{Subsect2} and we investigate the role of the topological charges for changing the chirality of fermions and also for creating stable zero and near-zero modes. In the last step, in section~\ref{Sect3}, the main points of our study are summarized. The results clearly show the role of topological charges for $\chi$SB. 
 
 \section{Plane vortex pairs, colorful regions and the Dirac modes}\label{Sect2}
 The configurations which we want to investigate are combinations of the thick plane vortices and colorful regions. The construction of the uni-color plane vortices have been explained in \cite{Jordan:2007ff}. The plane vortices extend along two coordinate axes and vortex thickness is added in a third direction and formulated with non-trivial links in the forth direction. The periodic boundary conditions are considered for the gauge fields and therefore vortices occur in pairs
of parallel sheets. We use plane vortices which extend in xy-plane, called xy-vortices. The t-links of the uni-color xy-vortices are nontrivial in one t-slice only and are elements in a U($1$) subgroup of SU($2$) and vary in $\sigma_3$-subgroup as
\begin{equation}\label{uni-color-plane}
U_4(x)=\exp\{\mathrm i\alpha(z)\sigma_3\}.
\end{equation}

The angle $\alpha$ is chosen as a linear function of z:
\begin{equation}
  \alpha_1(z) = \begin{cases}     2\pi \\ \pi\left[ 2-\frac{z-(z_1-d)}{2d}\right] \\ 
                           \pi \\ \pi\left[ 1-\frac{z-(z_2-d)}{2d}\right] \\ 
                          0 \end{cases} \ldots 
  \alpha_2(z) = \begin{cases}     0 & 0 < z \leq z_1-d, \\ 
                \frac{\pi}{2d}[z-(z_1-d)] & z_1-d < z \leq z_1+d, \\ 
                \pi & z_1+d < z \leq z_2-d, \\ 
                \pi\left[ 1-\frac{z-(z_2-d)}{2d}\right] & z_2-d < z \leq z_2+d, \\ 
                0 & z_2+d < z \leq N, \end{cases} \label{eq:alpha}
\end{equation}
where the vortex sheets have thickness $2d$ around $z_1$ and $z_2$. The gradient of the angle $\alpha$ determines the orientation of the vortex flux. The gradient of $\alpha_1$ ($\alpha_2$) at the two vortex sheets of a vortex pair points in the same (opposite) orientation and therefore their fluxes are (anti-)parallel and the total flux is $2\pi$ (zero). The vortex pairs with the same (opposite) vortex orientation are called parallel (anti-parallel) plane pair.

The topological charge of a field configuration obtains as 
\begin{equation}\label{eq:qlatq}
  Q=-\frac{1}{32\pi^2}\int d^4x \epsilon_{\mu\nu\rho\sigma} \mbox{tr}[{\mathcal F}_{\mu\nu}{\mathcal F}_{\rho\sigma}],
\end{equation}
where only regions with common presence of electric and magnetic fields of the same spatial directions contribute to the topological charge. Therefore, for the plane vortices which posses only non-trivial temporal links $U_4$, the gluonic
lattice topological charge is zero value.

 The effect of the configurations on fermions $\psi$ can be investigated by determining the low-lying eigenvectors and eigenvalues $|\lambda| \in [0,2/a]$ of the overlap Dirac operator \cite{narayanan:1993ss,narayanan:1994gw,neuberger:1997fp,edwards:1998yw}
\begin{eqnarray}\label{eq:ov_dirac}
D_{ov}=\frac{1}{a}\left[1+ \gamma_5 \frac{H}{|H|}\right]
\textrm{ with }H=\gamma_5 (a D_\mathrm{W}-m),\;
\end{eqnarray}
where $m$ denotes one species of single massless Dirac fermions on a lattice with lattice constant
$a$. The mass
parameter m is chosen with $m=+1$ and the lattice constant is puted $a=1$. The massless Wilson Dirac operator $D_\mathrm{W}$ on a lattice is
\begin{eqnarray}\label{WilsAct}
D_\mathrm{W}(x,y)=\frac{4}{a}\delta_{x,y}-
\frac{1}{2a}\sum_{\mu=\mp1}^{\pm4}(1-\gamma_\mu)\;U_\mu(x)\;\delta_{x+\hat\mu,y}.\;
\end{eqnarray}

The eigenvalues of the operator $D_{ov}$ as a Ginsparg-Wilson operator are on a circle in the complex plane. The absolute value $|\lambda|$ of the two complex conjugate eigenvalues of the overlap operator is simply written as $\lambda$. 

A general picture about $\chi$SB is that the topological charges can attract zero modes and produce a finite density of near-zero modes leading to $\chi$SB via the Banks-Casher relation \cite{banks:1979yr}.

In Ref.~\cite{hollwieser:2013xja}, for the overlap Dirac operator in the background of the single pairs of uni-color plane vortices, some zero and near-zero modes are calculated which can be removed by appropriate boundary conditions and therefore these modes are not stable. 

It is interesting what occurs for the non-stable zero and near-zero modes of single pairs of the uni-color plane vortices after adding colorful regions. We study the role of the topological charges for creating stable zero and near-zero modes.
 
\subsection{Combination of the uni-color plane and colorful spherical vortices}\label{Subsect1}

We combine the uni-color plane \cite{Jordan:2007ff,Hollwieser:2011uj} and colorful spherical vortices and analyze the response of fermions to this configurations. The construction of the spherical vortex has been
explained in~\cite{Jordan:2007ff,Hollwieser:2010mj,Schweigler:2012ae}. The spherical vortex represents a transition in the temporal direction between two pure gauge fields with the $SU(2)$ links as:
\begin{align}
\begin{split}
	U_{i} (x) &= \begin{cases}
 \left( g\left( \vec{r} + \hat{i} \right) \ g\left(\vec{r}\right)^{\dagger} \right)^{(t-1)/\Delta t}
 & \mbox{for} \quad 1 < t < 1 + \Delta t, \\
 g\left(\vec{r} + \hat{i}\right) \ g\left(\vec{r}\right)^{\dagger}
 & \mbox{for} \quad 1 + \Delta t \leq t \leq t_g, \\
 {1} & \mbox{else,}
\end{cases} , \\
	U_{4} (x) &= \begin{cases}
 g(\vec{r})^{\dagger}  & \mbox{for} \quad t = t_g, \\
 {1} &\mbox{else,}
\end{cases}
\label{links}
\end{split} 
\end{align}
where $\Delta t$ stands for the duration of the transition, $\vec{r}_0$ denotes the midpoint of the configuration and the function $g(\vec{r})$ is 
\begin{equation}\label{eq:g}
g(\vec{r}) = \cos  \left[  \alpha\left( \left| \vec{r} - \vec{r}_0 \right| \right) \right]  {1}
  - i \ \vec{e_r} \cdot \vec{\sigma} \ \sin \left[  \alpha\left( \left| \vec{r} - \vec{r}_0 \right| \right) \right],~~~
\vec{e_r} = \frac{\vec{r} - \vec{r}_0}{\left| \vec{r} - \vec{r}_0 \right|}.
\end{equation}
The angle $\alpha$ rises linearly as
\begin{gather}
\alpha(r) = \begin{cases} \pi & r < R-\frac{d}{2}, \\
                          \frac{\pi}{2}\left( 1-\frac{r-R}{\frac{d}{2}} \right) & R-\frac{d}{2} < r < R+\frac{d}{2}, \\
                          0 & R+\frac{d}{2} < r,
          \end{cases}\label{eq:alpha} \,
\end{gather}
where $R$ and $d$ denote the radius and thickness of spherical vortex respectively. 

\begin{figure}[h!]
\centering
a)\includegraphics[width=0.36\columnwidth]{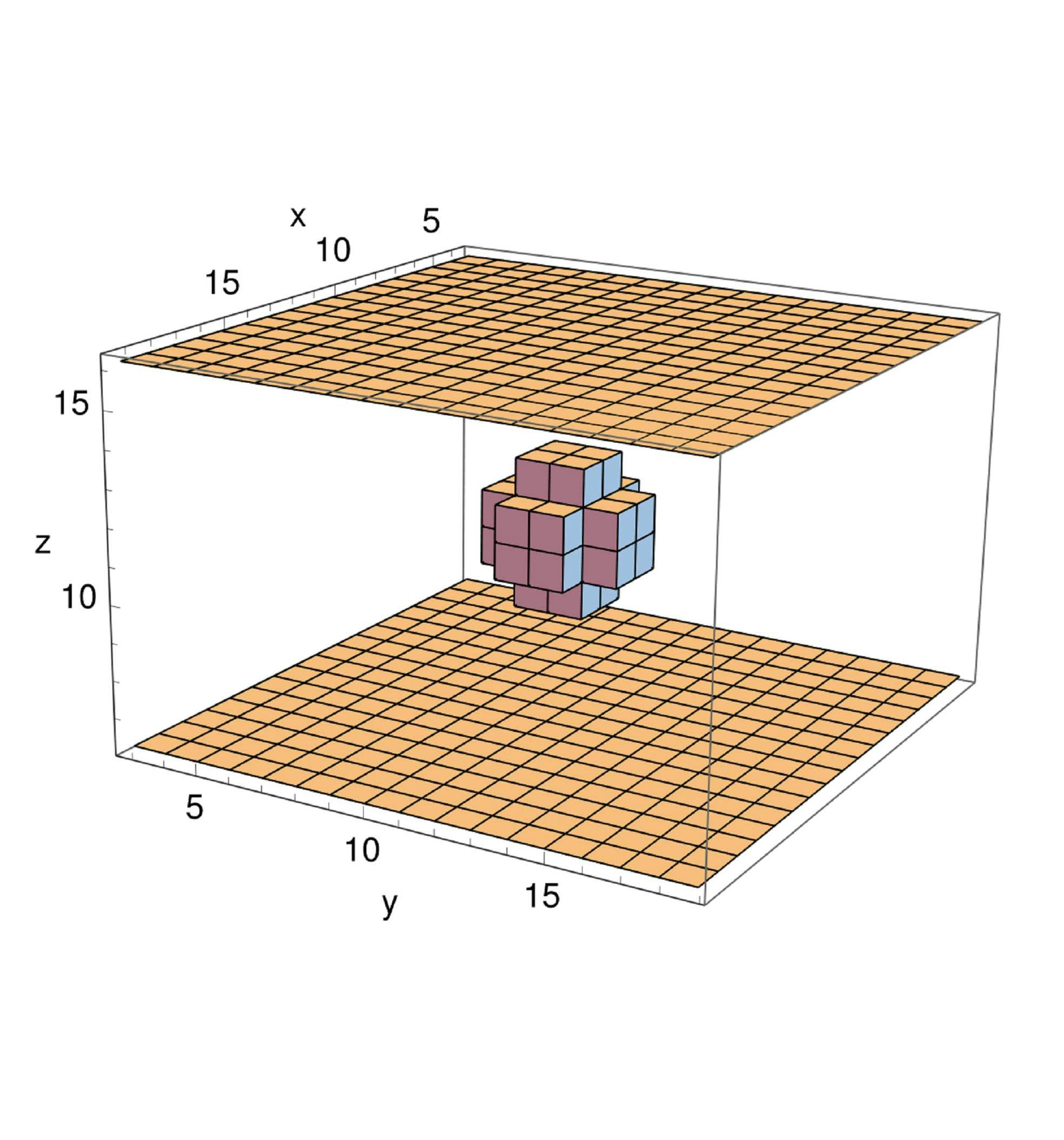}
b)\includegraphics[width=0.42\columnwidth]{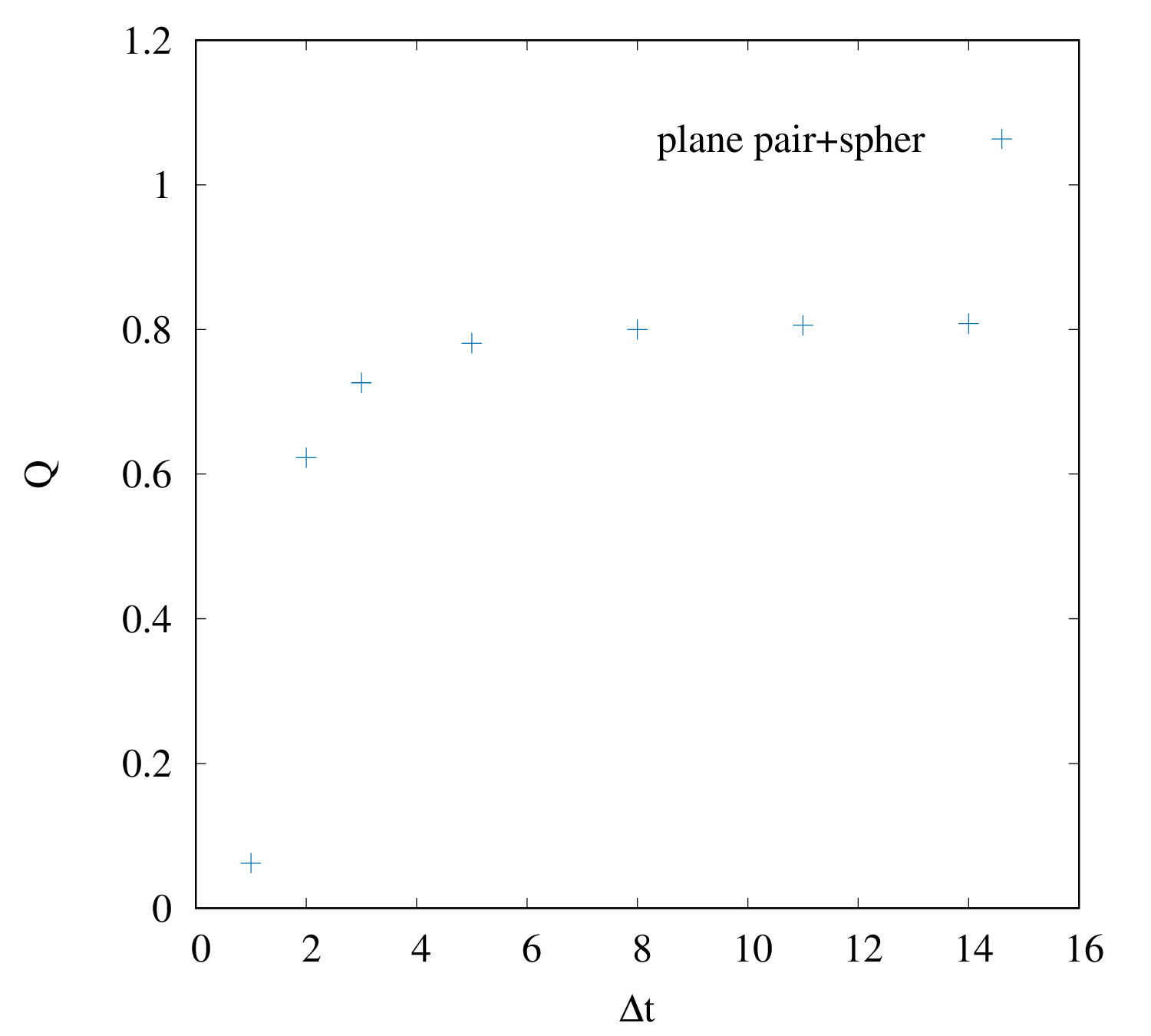}
\caption{a) Lattice representation of the combination of the plane and spherical vortices after maximal center gauge and center
projection. The spherical vortex locates in the middle of the configuration and the horizontal planes are the xy-vortices. b) The topological charge of the combination of the plane and spherical vortices on the lattice. The topological charge contribution of the uni-color plane vortices is zero while the one of the spherical vortex for slow transitions $\Delta t>1$ converges to near $+1$ and therefore the total topological charge of the combination increases to near $+1$. The calculations have been performed on
$16^4$-lattice.}
\label{fig:1}
\end{figure}

In Fig.~\ref{fig:1}a) we present a 3-dimensional view of the combination of the plane and spherical vortices on a $16^4$-lattice after maximal center gauge and center projection, leading to (thin) P-vortices. The position of the spherical vortex is in the middle of the configuration and the horizontal planes are the xy-vortices. In Fig.~\ref{fig:1}b), we show the topological charge of the combination as function of $\Delta t$ on a $16^4$-lattice. The gluonic topological charge contribution of the uni-color plane vortices is zero while the one of the spherical vortex converges to near $+1$ through increasing the value of $\Delta t$ and therefore the total topological charge of the combination increases to near $+1$.

\begin{figure}[h!]
\centering
a)\includegraphics[width=0.46\columnwidth]{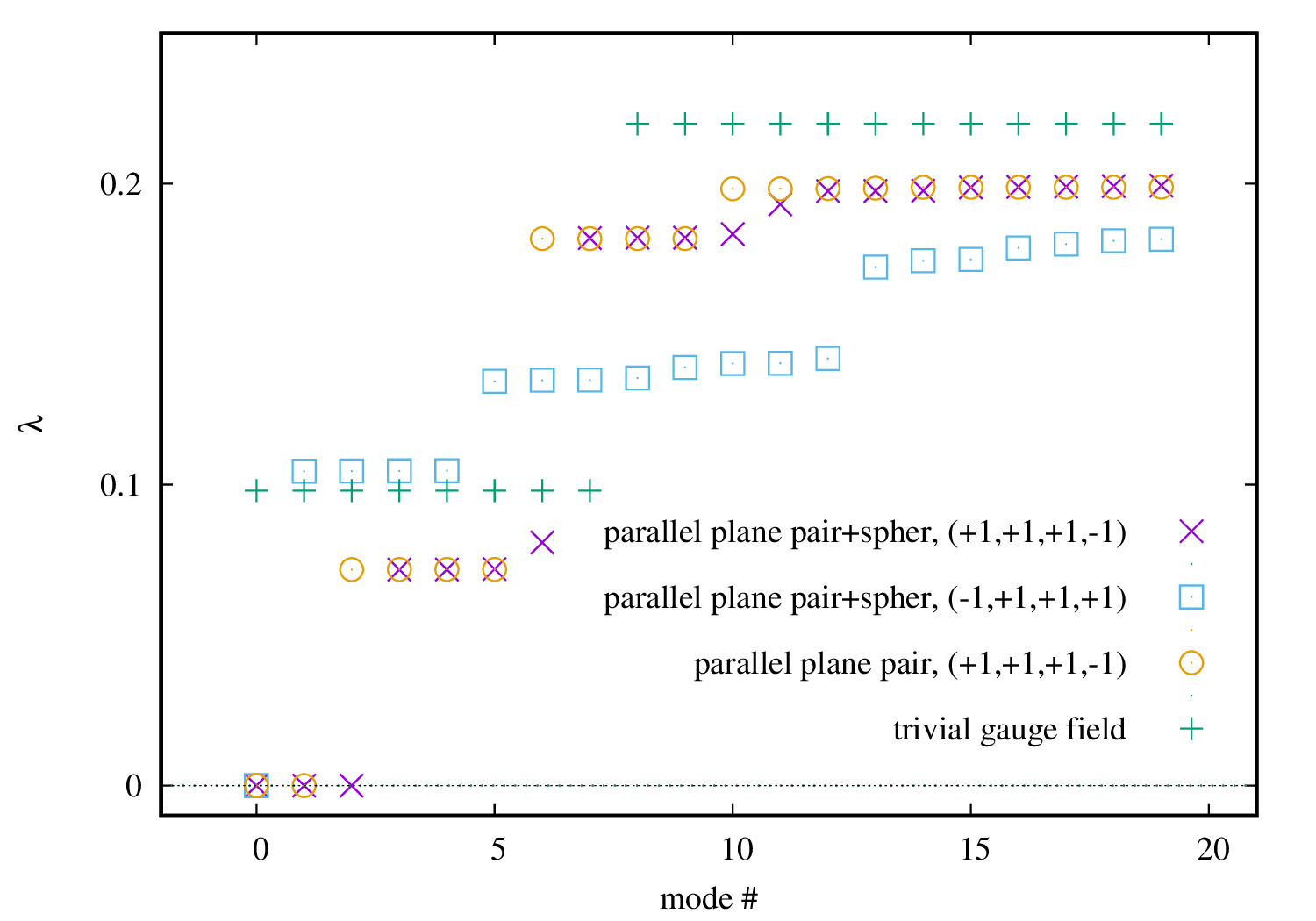}
b)\includegraphics[width=0.46\columnwidth]{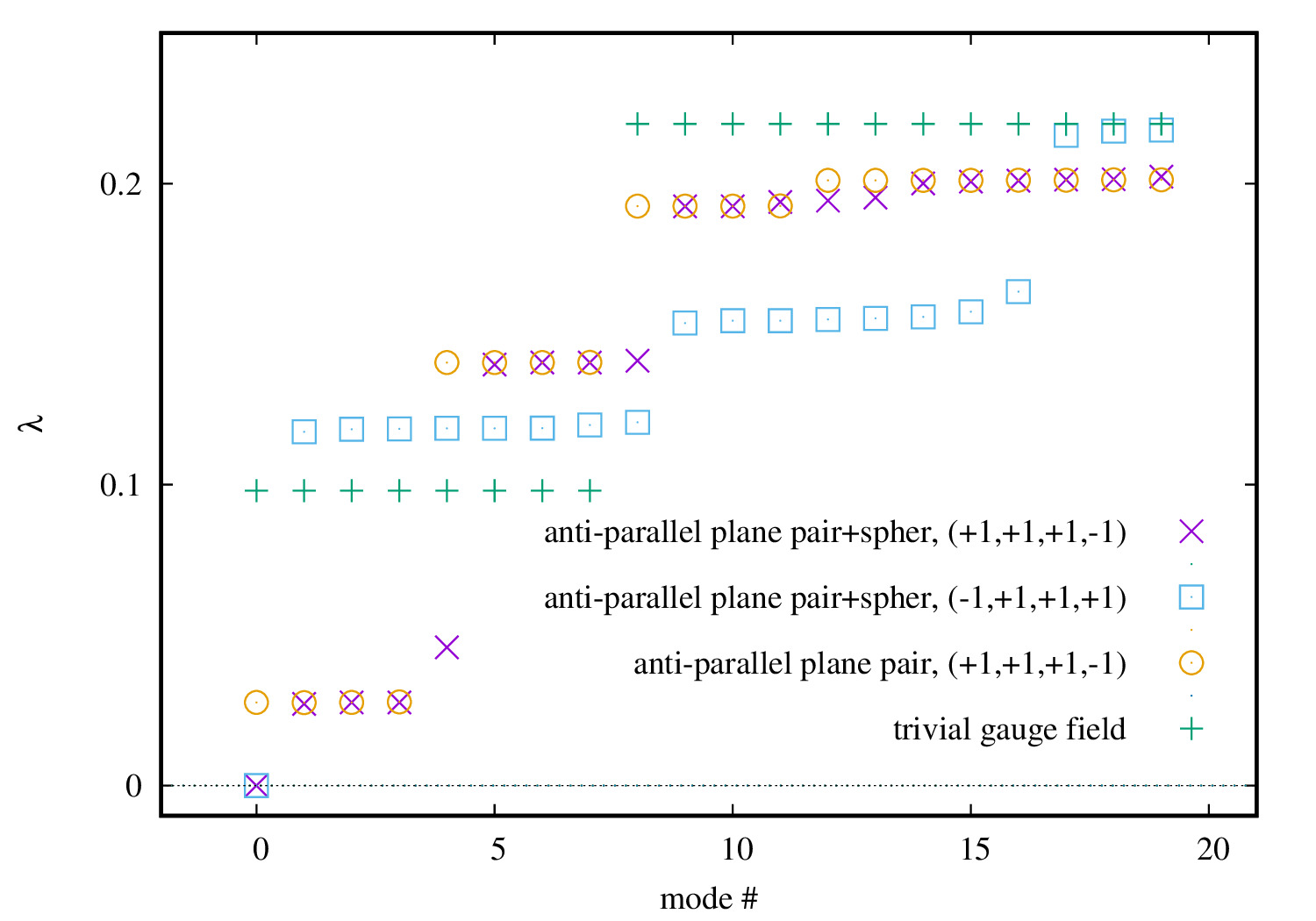}
\caption{a) The lowest overlap eigenvalues for the combination of the parallel plane and spherical vortices compared to the parallel plane vortex and free spectrums on $16^4$-lattices b) the same as a) but for anti-parallel ones. In the figures "+1"("-1") means (anti-)periodic boundary conditions in a direction(i.e., "the numbers in (+1,+1,+1,-1)" respectively means periodic boundary conditions in $x$, $y$, $z$ directions and anti-periodic boundary conditions in temporal direction).}
\label{fig:2}
\end{figure}

\begin{figure}[h!]
\centering
a)\includegraphics[width=0.28\columnwidth]{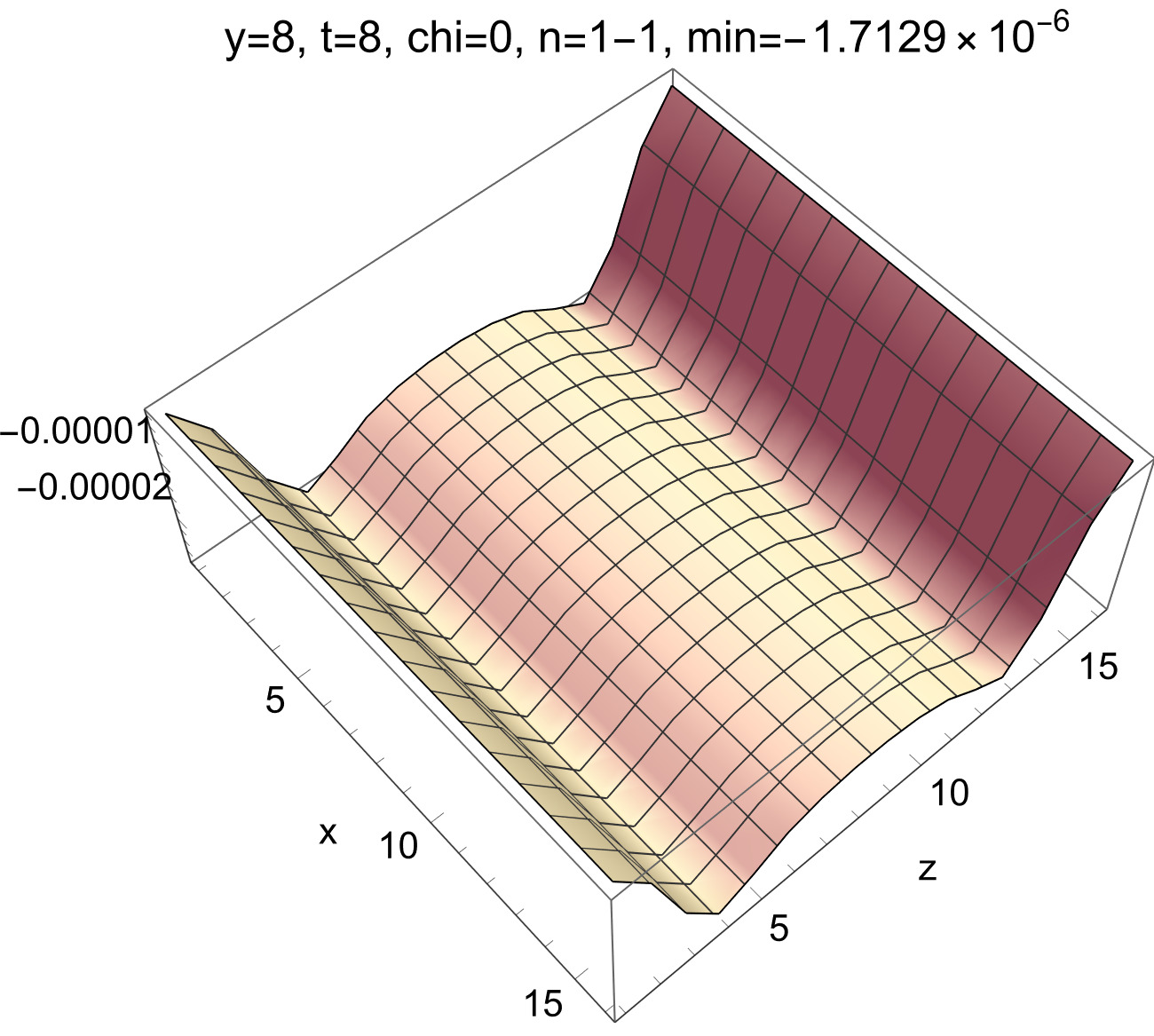}
b)\includegraphics[width=0.28\columnwidth]{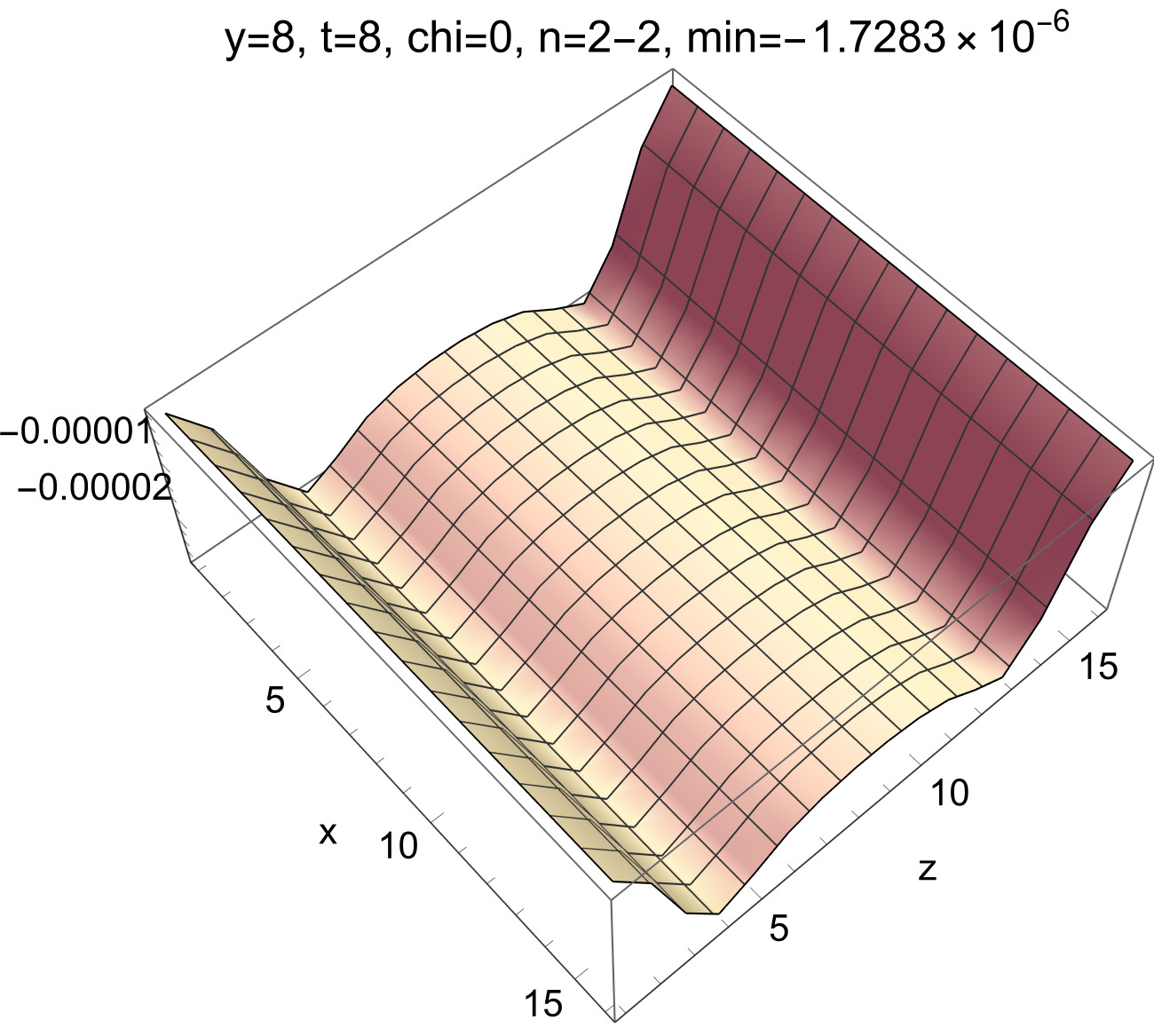}
\caption{Chiral densities of two zero modes with negative chirality for the parallel plane vortices in the x-z-planes. When there is not topological charge, the zero modes show nice plane wave oscillations.}
\label{fig:3}
\end{figure}

\begin{figure}[h!]
\centering
a)\includegraphics[width=0.28\columnwidth]{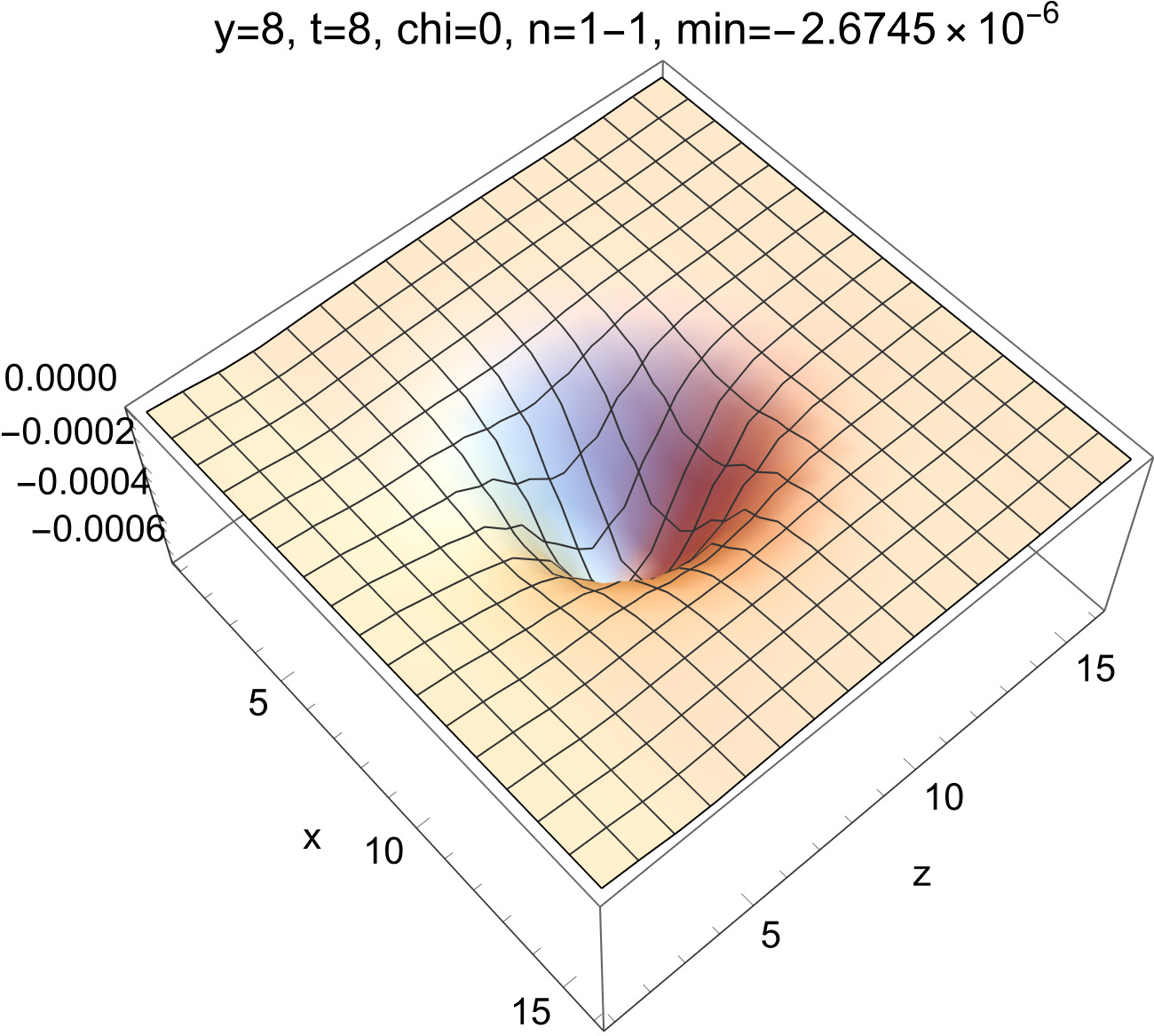}
b)\includegraphics[width=0.28\columnwidth]{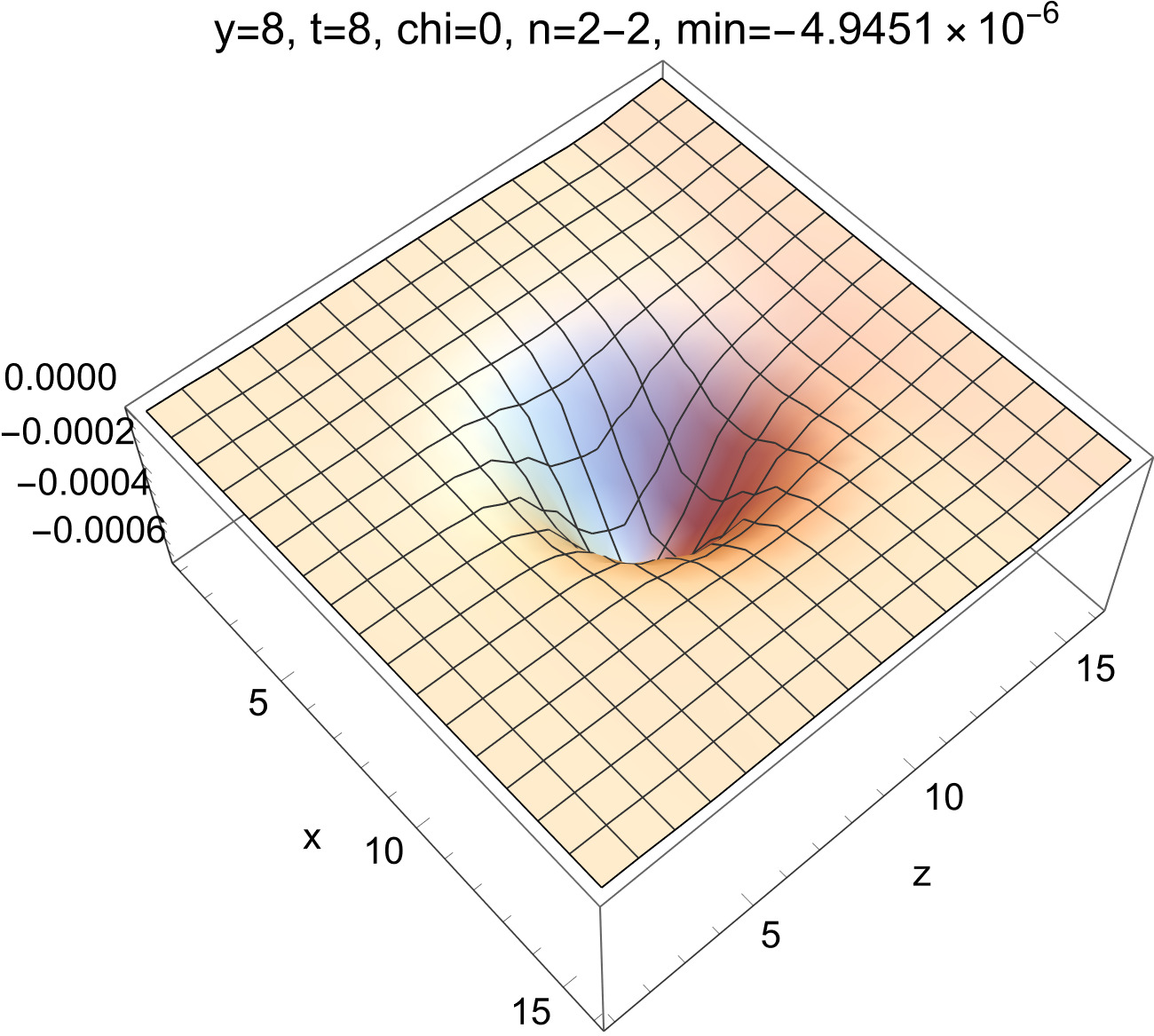}
c)\includegraphics[width=0.28\columnwidth]{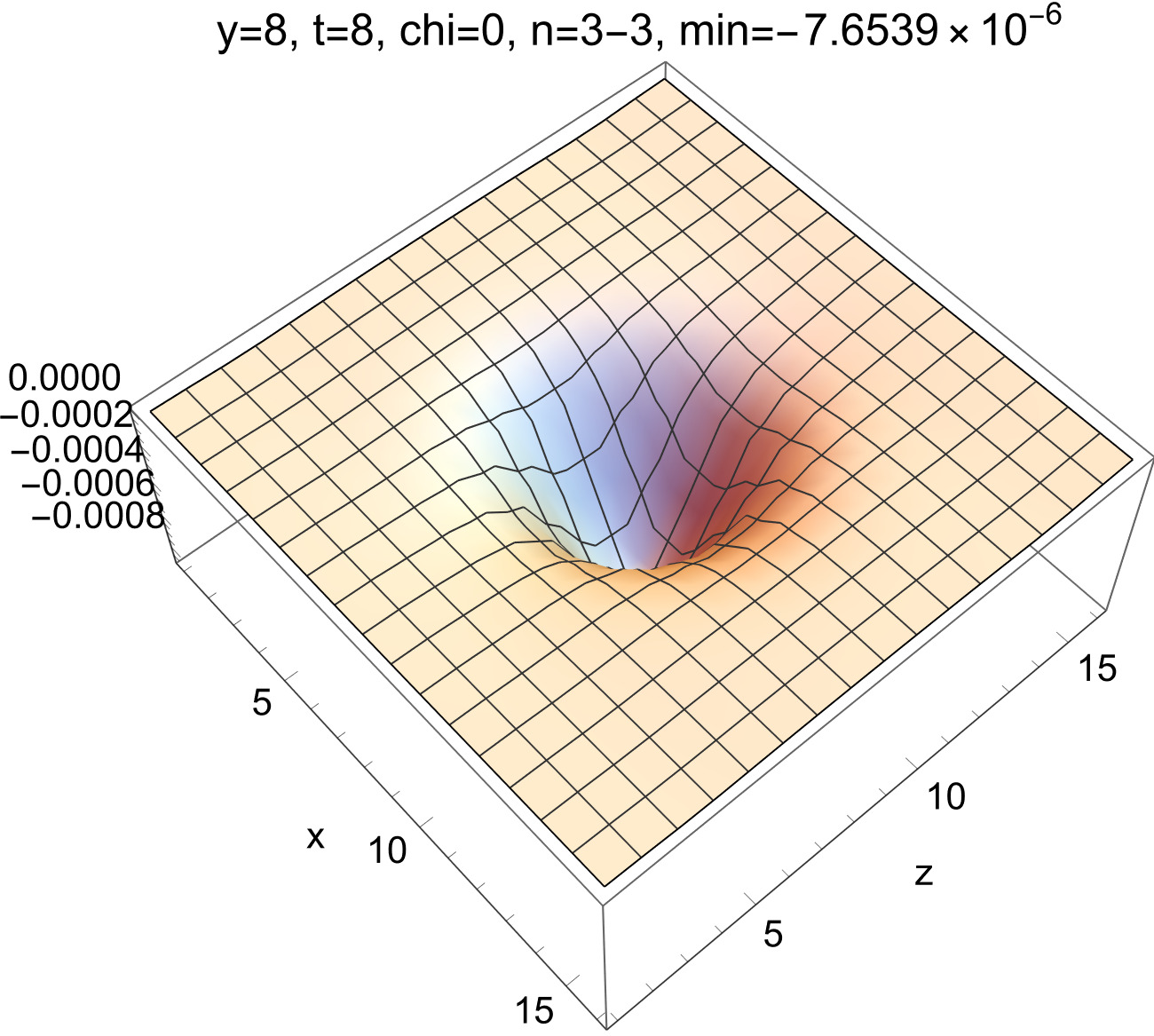}
\caption{Chiral densities of three zero modes with negative chirality for the combination of parallel plane and spherical vortices in the x-z-planes. One zero mode is attracted through the spherical vortex with topological charge $Q=+1$ and two zero modes are due to the parallel plane vortices. The chiral densities of all zero modes are localized around the spherical vortex. Therefore, two zero modes of the parallel plane vortices with plane wave oscillations in the background of the spherical vortex are localized in the topological charge region.}
\label{fig:4}
\end{figure}

In Fig.~\ref{fig:2}a), we show the eigenvalues of the combination of parallel plane and spherical vortices and compare them with parallel plane vortex and free spectrums on a $16^4$-lattice. Figure~\ref{fig:2}b) is the same as Fig.~\ref{fig:2}a) but for anti-parallel ones. 
For the fermions, first, we use anti-periodic boundary conditions in temporal direction and periodic boundary conditions in spatial directions. For the plane vortices, the vortex sheets with thickness $d=1.5$ are located around $z_1=3.5$ and $z_2=14.5$
respectively. For the spherical vortices, the center of the configuration with core radius $R=2$ and
thickness $d=1.5$ is located in the middle of the plane pair at $x=y=z=8.5$. As shown, for the parallel plane vortices, we get two left-handed zero modes. The chiral density of the eigenvector $\psi$, which gives the local chirality properties, is $\rho_5=\psi^\dagger\gamma_5\psi =\rho_+-\rho_-$ where $\rho_+$ and $\rho_-$ are right- and left-handed chiral densities \cite{hollwieser:2013xja}. In Fig.~\ref{fig:3}, we plot the chiral densities of these two left-handed zero modes which show nice plane wave oscillations. The plot titles of the chiral density give the $y$-
and $t$-coordinates of the $xz$-slices, "chi=0"
means we plot $\rho_5$, the number of plotted modes ("n=1-1" means we plot
$\rho\#1$, the chiral density of the lowest mode) and the minimal density in the plotted area ("min=...").
As shown in Fig.~\ref{fig:2}a), for the combination of parallel plane and spherical vortices, we get three left-handed zero modes. One zero mode is found because of the spherical vortex with topological charge $Q=+1$ and two zero modes are due to the parallel plane vortices. In Fig.~\ref{fig:4}, the chiral densities of these three zero modes are shown where they are localized around the spherical vortex. Therefore, two zero modes of parallel plane vortices does not show the plane wave oscillations and are attracted through the topological charge of the spherical vortex. 
As shown in Fig.~\ref{fig:2}a), although these two zero modes of parallel plane vortices in the background of spherical vortex are localized around the topological charge, they would be removed by appropriate boundary conditions. Using anti-periodic boundary conditions in $x$ direction and periodic boundary conditions in other directions for the fermions, for both parallel and anti-parallel configurations, we get one left-handed zero mode according to the total topological charge $Q=+1$ which agrees with the the Atiyah-Singer index, $\mathrm{ind}D[A]=n_--n_+=Q$, $n_-$ and $n_+$ denote the numbers of left- and right-handed zero modes~\cite{atiyah:1971rm,brown:1977bj,adams:2000rn}.

As a result, the topological charge can not stabilize the zero modes of the uni-color plane pair. 

In addition, we get four low-lying modes for the combination of the plane and spherical vortices for both parallel and anti-parallel. These four low-lying modes are smaller eigenvalues than the ones of the lowest eigenvectors for the trivial gauge field and one can identify them as near-zero modes showing plane wave oscillations similar to those of parallel and anti-parallel plane vortices. As expected, these near-zero modes are removed by changing the boundary conditions.

\subsection{Plane vortex pairs with two colorful vortices}\label{Subsect2}
Now, we investigate the interaction of topological charges through plane vortex pairs with two colorful vortices. In the previous section, the zero and near-zero modes of uni-color plane vortices in the background of topological charge of the spherical vortex can not persist through changing the boundary conditions. We investigate the zero modes attracted by topological charge and show how colorful plane vortices
form near-zero modes from zero modes through interactions. In Ref.~\cite{hollwieser:2013xja}, the near-zero mode is calculated when there is a time distance between spherical vortex and anti-vortex. Now, we investigate the low-lying modes for plane vortex pair with two colorful vortices where there is a distance in spatial direction between two colorful plane vortices. The construction of the plane vortices with one colorful plane has been introduced in \cite{HosseiniNejad:2015oeu}. For constructing the plane vortex pairs with two colorful vortices, the links are given by Eq.~(\ref{links}) but the function $g(\vec{r})$ is:
\begin{equation}\label{eq:sphv}
g(\vec{r}) =
\begin{cases}
\mathrm e^{\mathrm -i\alpha(z)\vec n\cdot\vec\sigma}\quad&\mathrm{for}\quad
z_1-d\leq z\leq z_1+d\quad\mathrm{and}\quad 0\leq\rho\leq R,
\\\mathrm e^{\mathrm -i\alpha(z)\vec n\cdot\vec\sigma}\quad&\mathrm{for}\quad
z_2-d\leq z\leq z_2+d\quad\mathrm{and}\quad 0\leq\rho\leq R,
\\\mathrm e^{\mathrm -i \alpha(z)\sigma_3}  & \mathrm{else},\end{cases}
\end{equation}
where
\begin{equation}\begin{aligned}\label{DefColVort}
&\vec n\cdot\vec\sigma=\sigma_1\,\sin\theta(\rho)\cos\phi
+\sigma_2\,\sin\theta(\rho)\sin\phi+\sigma_3\,\cos\theta(\rho),\\
&\theta(\rho)=\pi(1-\frac{\rho}{R}),\quad \rho=\sqrt{(x-x_0)^2+(y-y_0)^2},\\
&\phi=\arctan_2\frac{y-y_0}{x-x_0}\;\in\,[0,2\pi).
\end{aligned}\end{equation}
Therefore, one colorful cylindrical region is located in the first plane vortex around $z_1$ and another one in the second plane vortex around $z_2$. The center of the colorful cylindrical region in both vortex sheets with radius $R$ is located in the $xy$-plane at $(x_0,y_0)$.

\begin{figure}[h!]
\centering
a)\includegraphics[width=0.28\columnwidth]{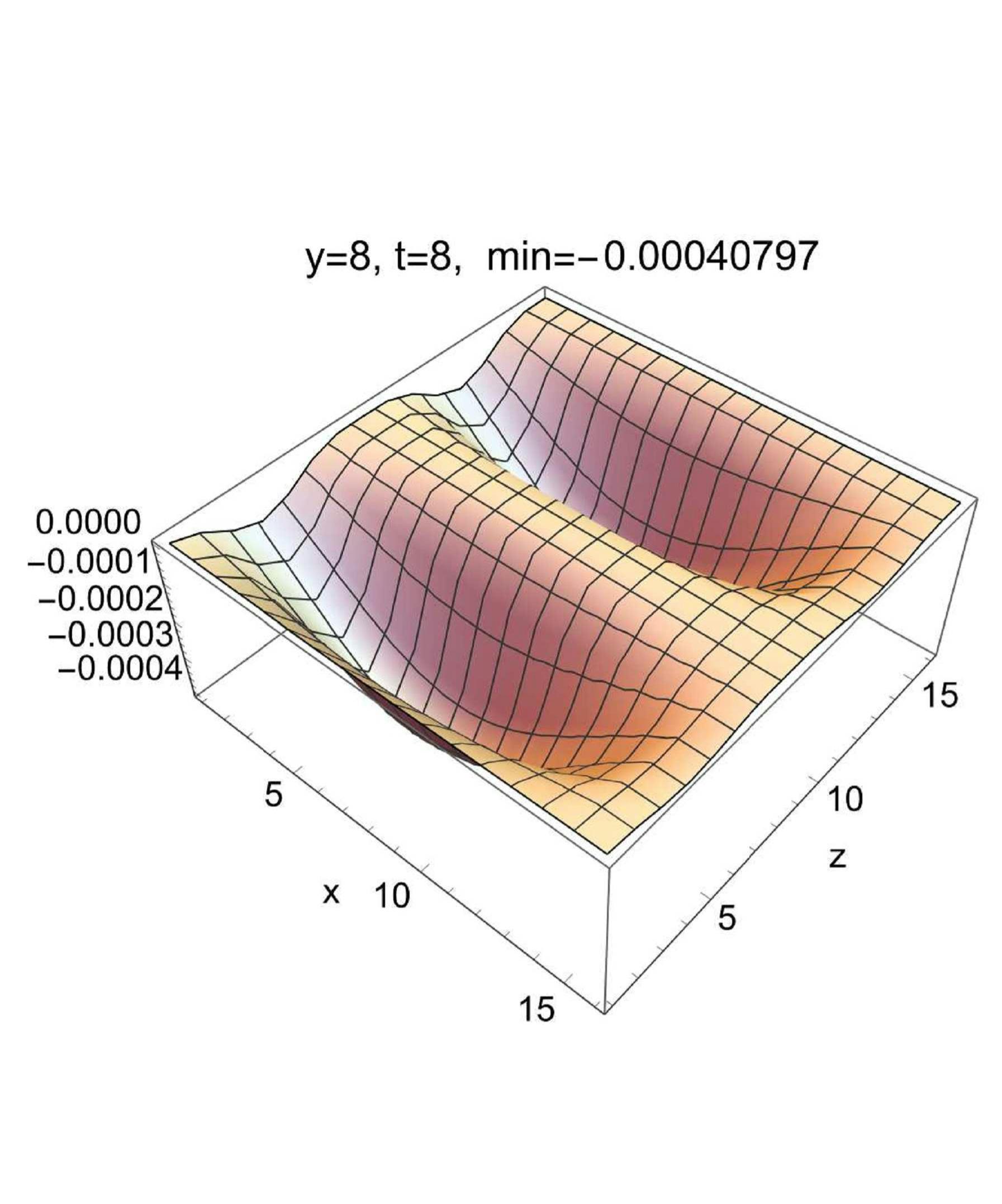}
b)\includegraphics[width=0.38\columnwidth]{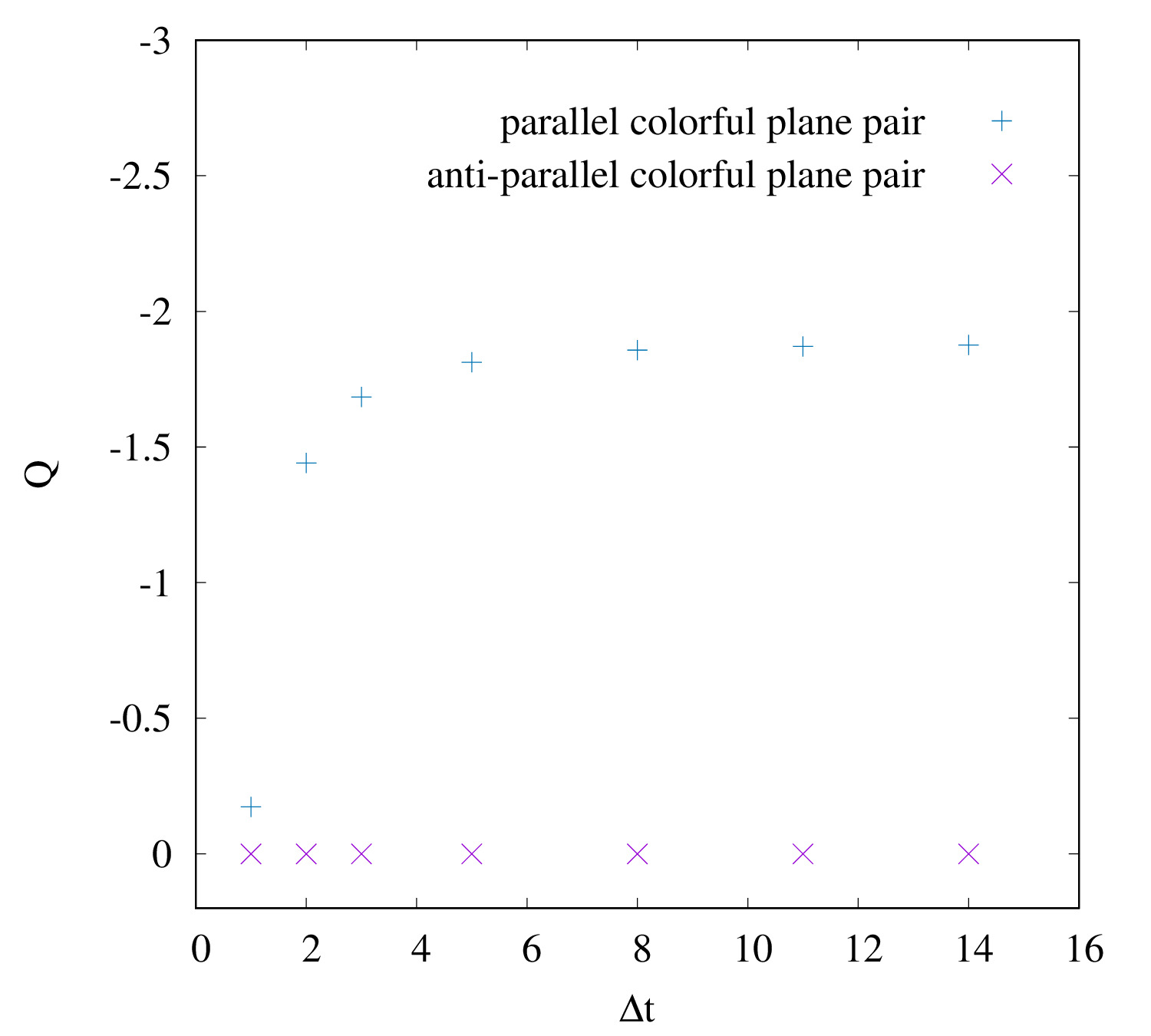}
c)\includegraphics[width=0.28\columnwidth]{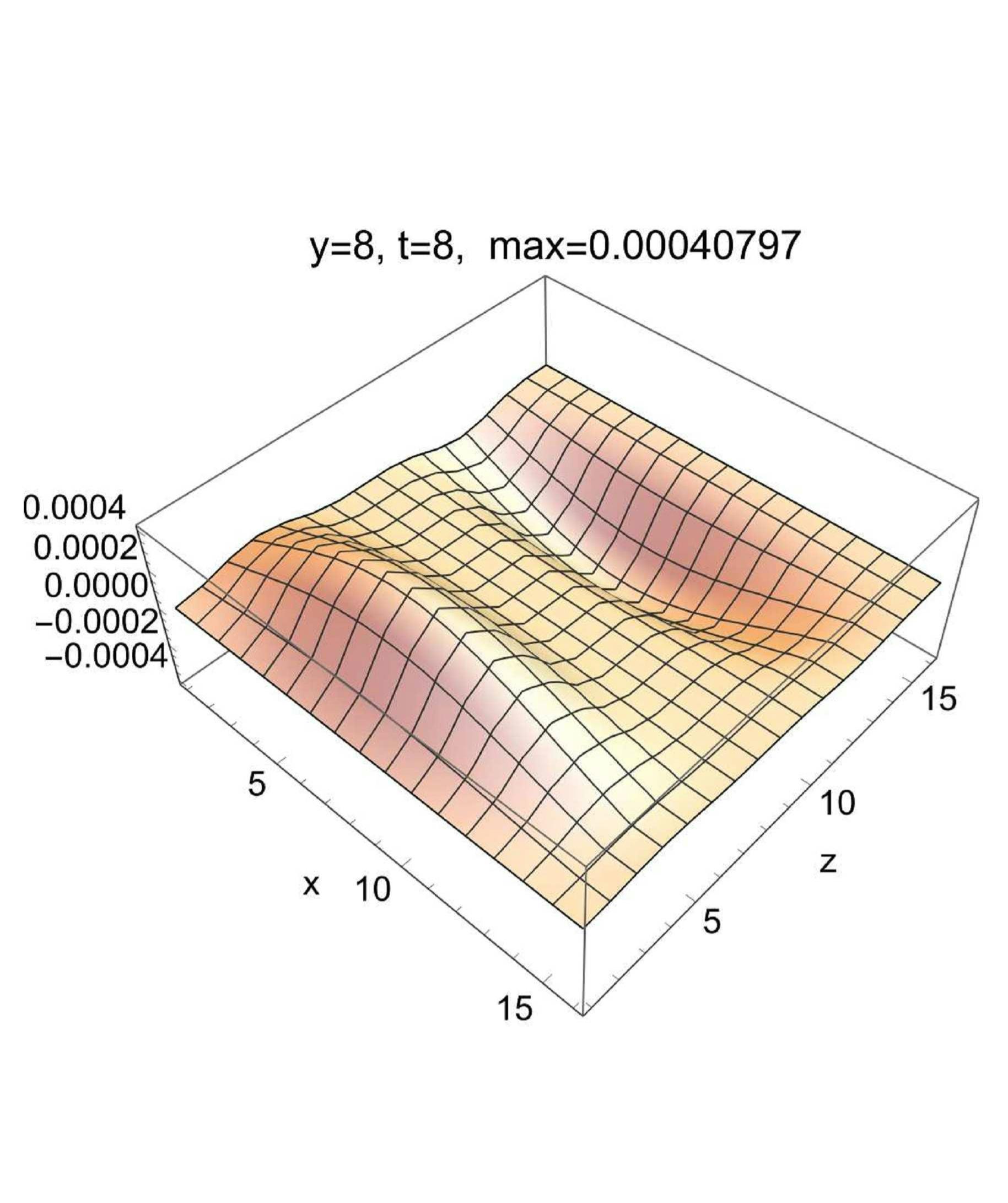}
\caption{The topological charges and their densities for the plane vortex pair with two colorful vortices. For the colorful plane vortices, the two vortex sheets with thickness $d=3$ are located around $z_1= 4.5$ and $z_2 = 12.5$ where the center of the colorful region with radius $R=8$ in both vortex sheets is located in the $xy$ plane at $x_0=y_0=8$. The topological charge for slow transitions converges to $Q=-2$ for the parallel colorful plane vortices. We get $Q(\Delta t)=0$ for all $\Delta t$ for the anti-parallel colorful plane vortices. The topological charge density of the $Q=-2$ ($Q=0$) configuration in the x-z-planes is given in a) (c)) at $y=t=8$, the middle of smoothed colorful regions. Each lump in the topological charge densities contributes $Q=+1$ or $Q=-1$.}
\label{fig:5}
\end{figure}

In Fig.~\ref{fig:5} we show the topological charges as well as their densities for the colorful plane vortices on a $16^4$-lattice. For colorful plane vortices, the two colorful vortex sheets with thickness $d=3$ are located around $z_1= 4.5$ and $z_2 = 12.5$ respectively. The center of the colorful region in both vortex sheets with radius $R=8$ is located in the $xy$ plane at $x_0=y_0=8$. The topological charge of the parallel colorful plane vortices for slow transitions ($\Delta t>1$) converges to near $-2$ and there are two localized lumps of topological charge $Q=-1$ in the topological charge density as shown in Fig.~\ref{fig:5}a). The lumps of topological charges are around the colorful regions of two vortex sheets and therefore, the topological charge contribution of each plane vortex is $-1$. For the anti-parallel colorful plane vortices, we get $Q(\Delta t)=0$ for all $\Delta t$ and there are two lumps of topological charge $Q=+1$ and $Q=-1$, the first (second) lump around the colorful region of the first (second) vortex sheet, shown in Fig.~\ref{fig:5}c).

\begin{figure}[h!]
\centering
a)\includegraphics[width=0.46\columnwidth]{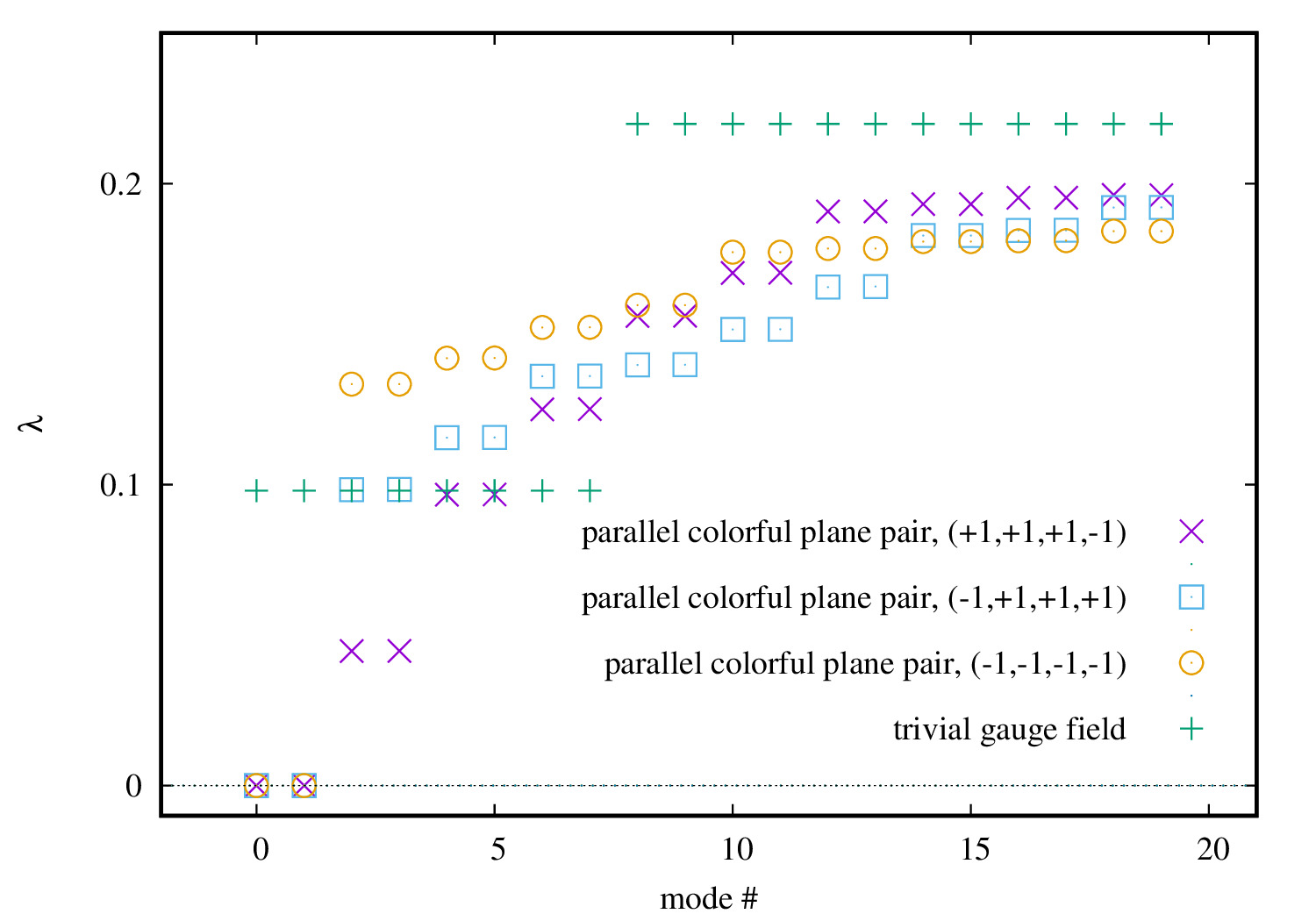}
b)\includegraphics[width=0.46\columnwidth]{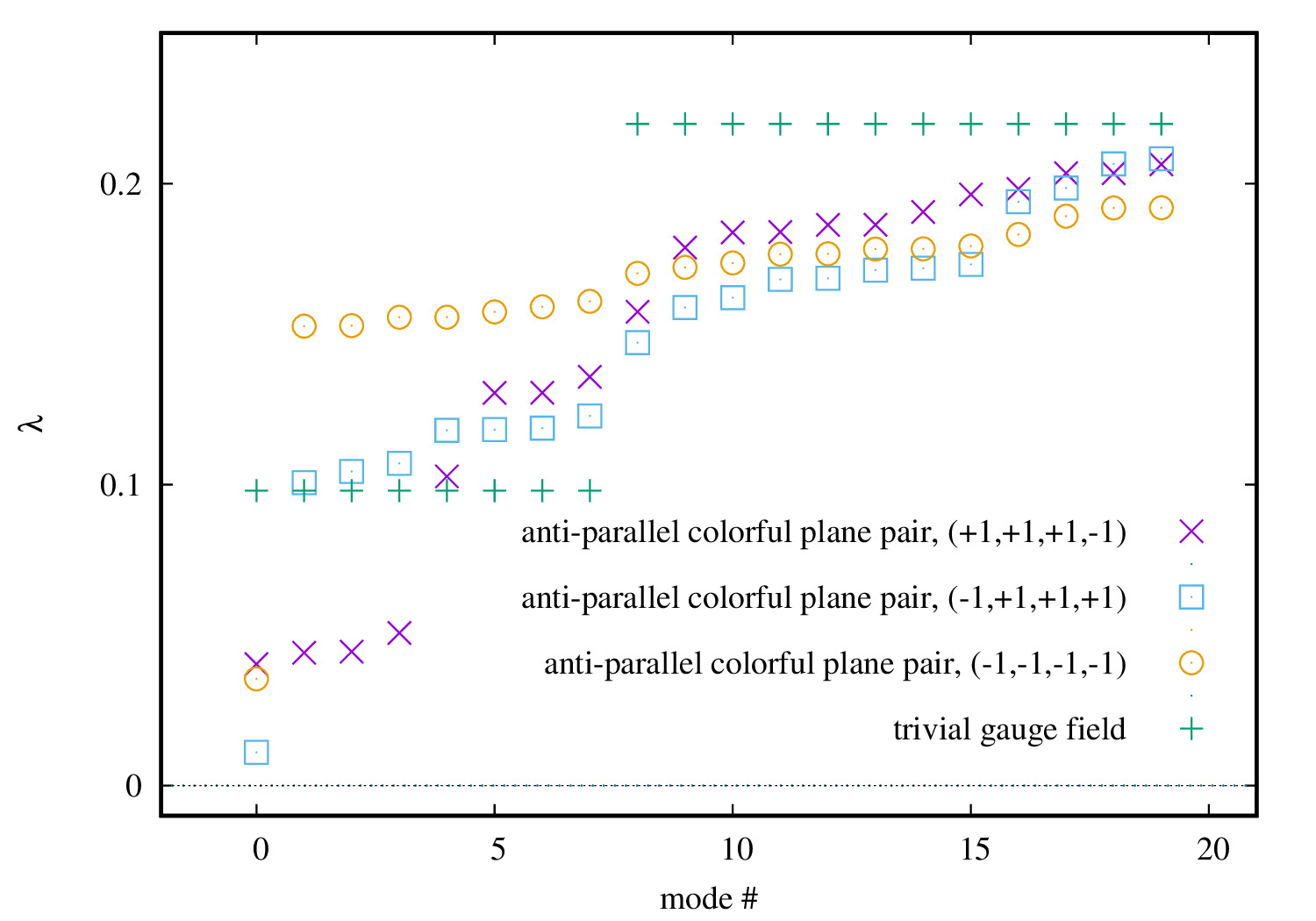}
\caption{a) The lowest overlap eigenvalues for the parallel colorful plane vortices compared
with those of the free Dirac operator on $16^4$-lattices b) the same as a) but for the anti-parallel colorful plane vortices. Two zero modes of the parallel colorful plane vortices and one near-zero mode of the anti-parallel colorful plane vortices persist regardless of boundary conditions even if we impose anti-periodic boundary conditions in all directions.}
\label{fig:6}
\end{figure}

In Fig.~\ref{fig:6}a), we plot the eigenvalues of the parallel colorful plane vortices and compare them with those of the free overlap Dirac operator on a $16^4$-lattice. Figure~\ref{fig:6}b) is the same as Fig.~\ref{fig:6}a) but for the anti-parallel colorful plane vortices. The parameters of the configurations
are the same as those in Fig.~\ref{fig:5}.

\begin{figure}[h!]
\centering
a)\includegraphics[width=0.28\columnwidth]{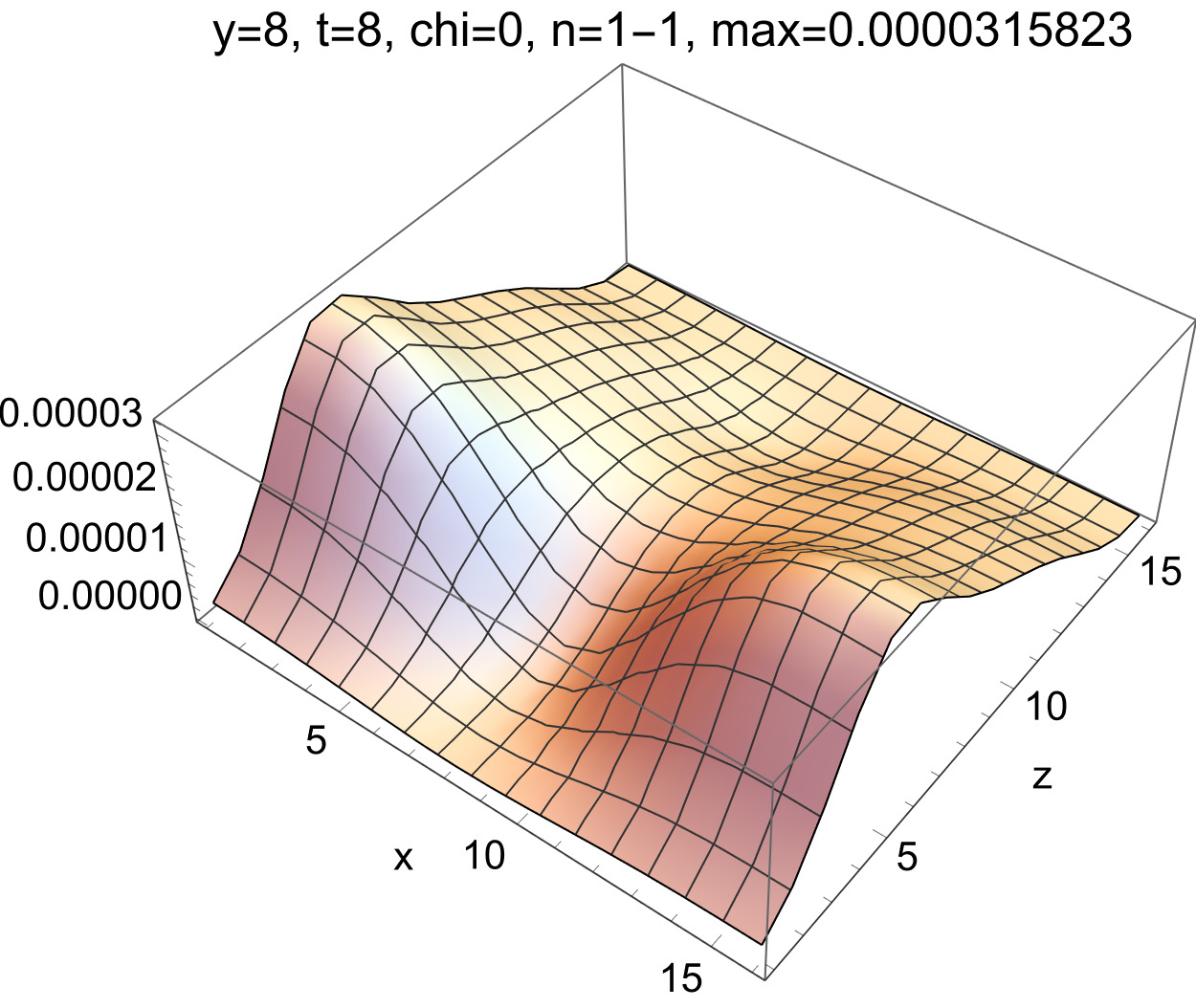}
b)\includegraphics[width=0.28\columnwidth]{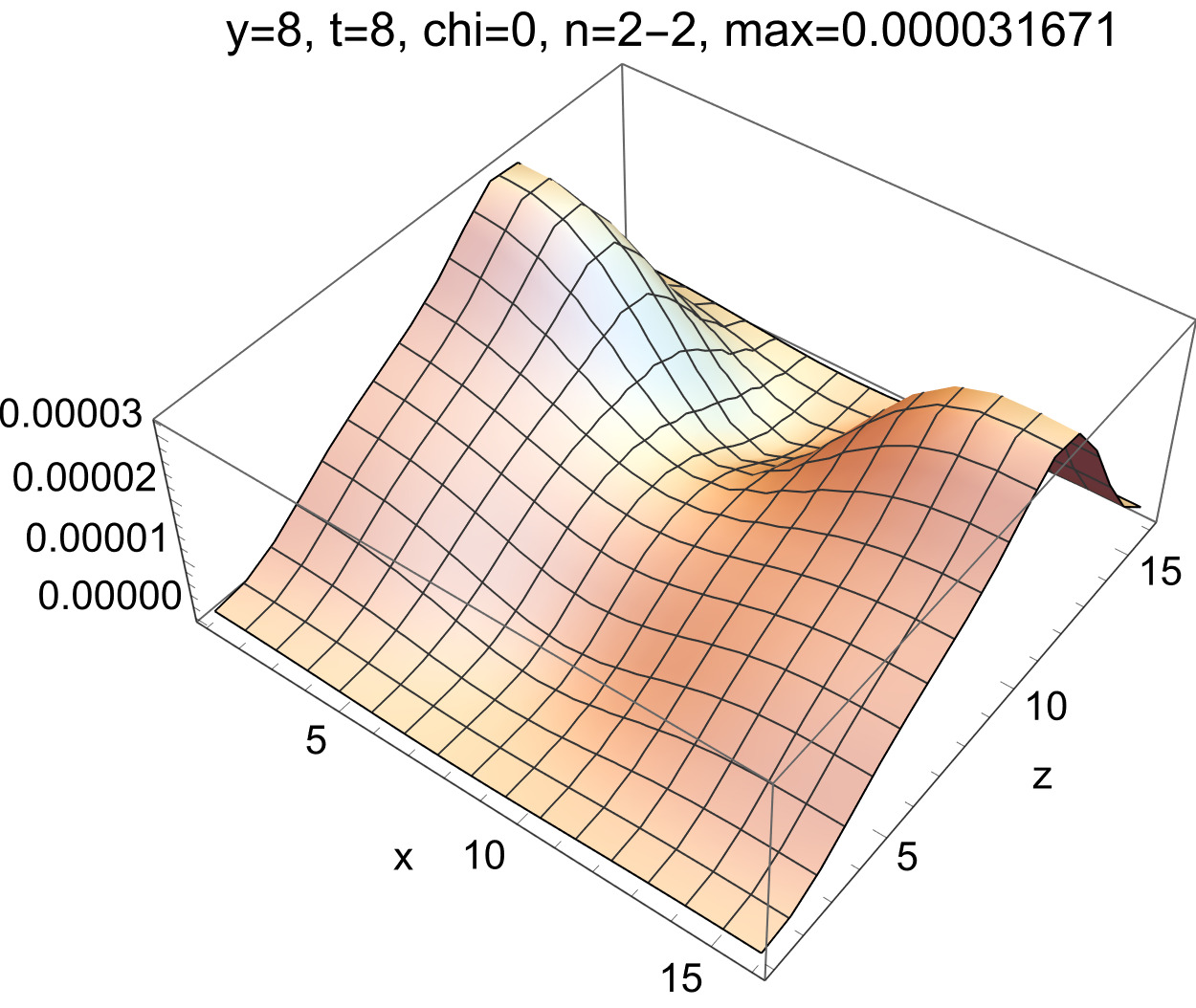}
c)\includegraphics[width=0.28\columnwidth]{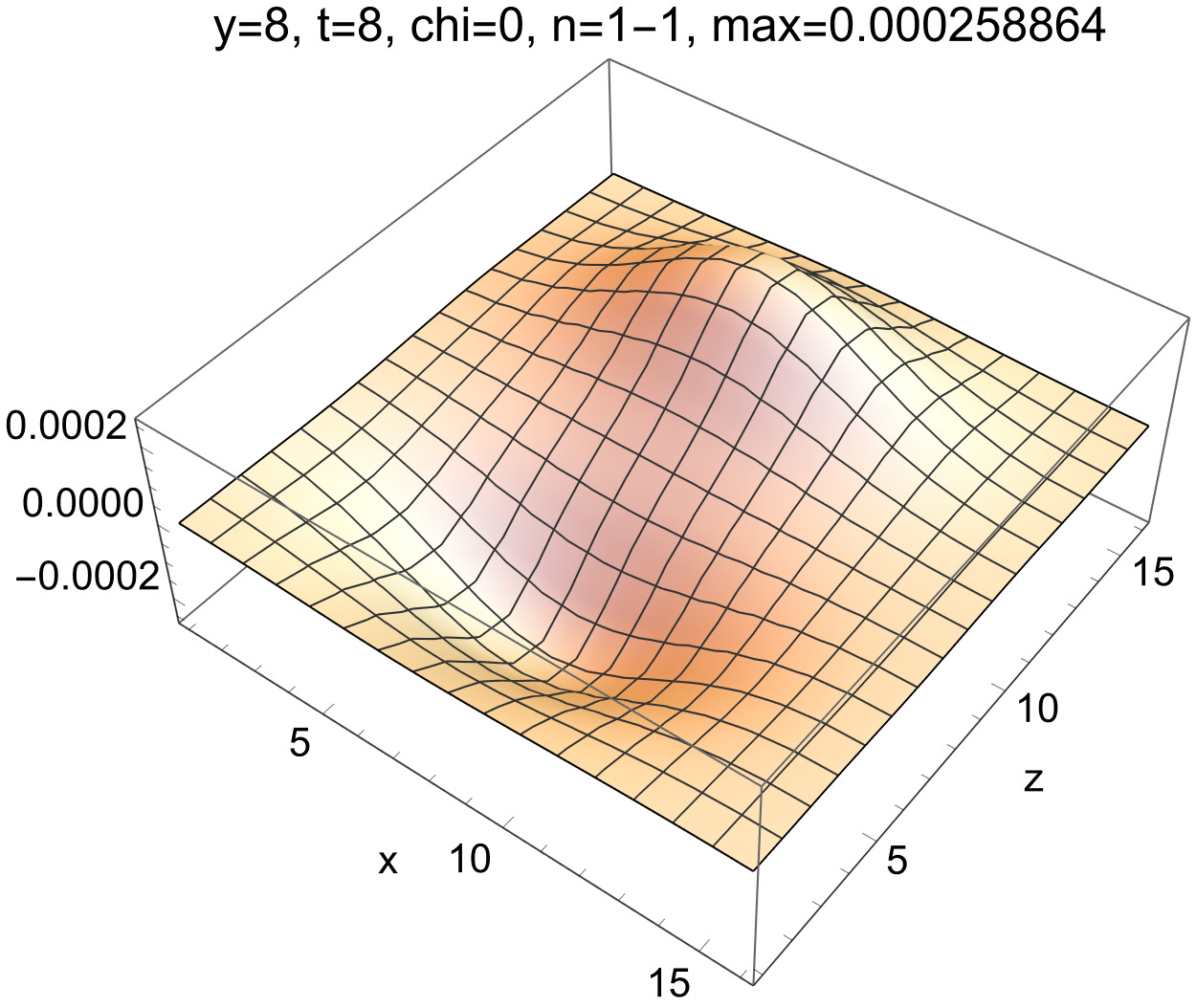}
\caption{Chiral densities corresponding to the diagrams of the topological charge density given in Fig.~\ref{fig:5} in the x-z-planes a) and b) for the two zero modes with positive chirality for the parallel colorful plane vortices. Each zero mode is localized in one colorful vortex sheet. c) for the near-zero mode of the anti-parallel colorful plane vortices, localized in the colorful regions of two colorful vortex sheets with opposite topological charge contributions. }
\label{fig:7}
\end{figure}

As shown, for the parallel colorful plane vortices, we get two right-handed zero modes, according to the total topological charge $Q=-2$ and no zero mode for the anti-parallel colorful plane vortices with total topological charge $Q=0$ as expected from the index theorem. We want to emphasize that these two zero modes persist regardless of boundary conditions even if we impose anti-periodic boundary conditions in all directions, as shown in Fig.~\ref{fig:6}a). In Figs.~\ref{fig:7}a) and b), we show the chiral densities of two zero modes of the $Q=-2$ configuration with positive chirality in the x-z-planes. Each vortex sheet of the $Q=-2$ configuration with topological charge contribution $Q=-1$, as shown in Fig.~\ref{fig:5}a), attracts one zero mode with positive chirality. 

Therefore two left-handed zero modes (negative chiralities) of parallel plane vortices with plane wave oscillations, as shown in Figs.~\ref{fig:3} after adding topological charges to both vortex sheets become two localized right-handed zero modes. It may show the role of the topological charge for changing the handedness of fermions which is the characteristic property for $\chi$SB. 

As a result, the topological charge may be able to change the chirality of the zero modes. 

In addition, using anti-periodic boundary conditions in temporal direction and periodic boundary conditions
in spatial directions for the fermions, we get some low-lying modes for the colorful plane vortices for both parallel and anti-parallel. These low-lying modes are smaller eigenvalues than the ones of the lowest eigenvectors for the trivial gauge field and one can identify them as near-zero modes. However using anti-periodic boundary conditions in $x$ direction and periodic boundary conditions
in other directions for the fermions, only one near-zero mode for the anti-parallel colorful plane vortices persist and there is no near-zero modes for the parallel colorful plane vortices. Even if we impose anti-periodic boundary conditions in all directions, this near-zero mode of the anti-parallel colorful plane vortices does not change, as shown in Fig.~\ref{fig:6}b). In Fig.~\ref{fig:7}c), we show the chiral density of the near-zero mode of the $Q=0$ configuration in the x-z-planes. The near-zero mode is localized in the colorful regions of two vortex sheets of the $Q=0$ configuration with opposite topological charge contributions $Q=+1$ and $Q=-1$, as shown in Fig.~\ref{fig:5}c).

Therefore, anti-parallel plane vortex pairs with two colorful vortices may contribute to a
finite density of near-zero modes leading to $\chi$SB via the Banks-Casher relation.

\section{Conclusion}\label{Sect3}
We investigate combinations of the plane vortex pairs with colorful regions. The influence of the topological charges is analyzed on the non-stable zero and near-zero modes of the overlap Dirac operator in the background of uni-color plane vortices. For the parallel uni-color plane vortices, there are two left-handed zero modes which their chiral densities show nice plane wave oscillations. However these two zero modes can be removed by the appropriate boundary conditions. 

Combining the parallel uni-color plane vortices with the spherical vortices with topological charge $Q=+1$, we get three zero modes where one zero mode is found because of the spherical vortex and two zero modes are due to the parallel plane vortices. The chiral densities of these zero modes are localized around the spherical vortex. Therefore, two zero modes of parallel plane vortices does not show the plane wave oscillations and are attracted by the topological charge of the spherical vortex. Although two zero modes of the parallel plane vortices localize around the topological charge, they can not persist by changing the boundary conditions. Therefore, it seems that the topological charges can not stabilize the zero modes of the uni-color plane pair. For the parallel (anti-parallel) uni-color plane vortices with the spherical vortices, we get one stable left-handed zero mode according to the total topological charge $Q=+1$ which agrees with the the Atiyah-Singer index. In addition, we get four low-lying modes identified as near-zero modes for the combination of the plane and spherical vortices for both parallel and anti-parallel. However the chiral densities of these near-zero modes show the plane wave oscillations similar to those of the parallel and anti-parallel plane vortices and they are removed by changing the boundary conditions.

Then we study the creation of near-zero modes by topological charges. We construct plane vortex pairs with two colorful vortices and show how colorful plane vortices form near-zero modes from zero modes through interactions. For the parallel (anti-parallel) colorful plane vortices, we get $Q=-2$($Q=0$) for the topological charge and each $+1$ or $-1$ topological charge contribution is a lump around colorful vortex sheets in the topological charge densities. For the parallel colorful plane vortices, we get two right-handed zero modes, according to the total topological charge $Q=-2$ and no zero mode for the anti-parallel colorful plane vortices with total topological charge $Q=0$ as expected from the index theorem. We want to emphasize that these two zero modes persist regardless of boundary conditions even if we impose anti-periodic boundary conditions in all directions. Therefore two left-handed zero modes (negative chiralities) of parallel plane vortices with plane wave oscillations after adding topological charges to both vortex sheets become two localized right-handed zero modes (positive chiralities). It may show the role of the topological charge for changing the handedness of fermions which is the characteristic property for spontaneous chiral symmetry breaking. As a result, the topological charge may be able to change the chirality of the zero modes. In addition, we get some near-zero modes for the colorful plane vortices for both parallel and anti-parallel. However changing the boundary conditions for the fermions, only one near-zero mode for the anti-parallel colorful plane vortices persist. Even if we impose anti-periodic boundary conditions in all directions, this near-zero mode does not change. This near-zero mode is localized in the colorful regions of two vortex sheets of the $Q=0$ configuration with opposite topological charge contributions $Q=+1$ and $Q=-1$.

Therefore, anti-parallel plane vortex pairs with two colorful vortices are able to create a finite
density of near-zero modes leading to $\chi$SB via the Banks-Casher relation.

\section{\boldmath Acknowledgments}

The author is grateful to the research council
of Semnan University and Iran National Science Foundation (INSF) for supporting this study. This work is based upon research funded by Iran National Scinence Foundation (INSF) under project No.4013593.

\bibliographystyle{unsrt}
\bibliography{chiral}

\end{document}